\documentclass[12pt,preprint]{aastex}
\begin{document}
\title{The Intrinsic Alignment of Dark Halo Substructures}
\author{\sc Jounghun Lee\altaffilmark{1,2}, Xi Kang\altaffilmark{3}, and
Yipeng Jing\altaffilmark{3}}
\altaffiltext{1}{School of Physics, Korea Institute for Advanced Study,
Seoul 207-43, Korea ; jounghun@newton.kias.re.kr}
\altaffiltext{2}{Astronomy Program, School of Earth and Environmental
Sciences, Seoul National University, Seoul 151-742 , Korea}
\altaffiltext{3}{Shanghai Astronomical Observatory; the Partner Group of MPA,
Nandan Road 80, Shanghai 200030, China ; kangx@shao.ac.cn}
\begin{abstract}
We investigate the intrinsic alignments of dark halo substructures
with their host halo major-axis orientations both analytically and
numerically. Analytically, we derive the probability density
distribution of the angles between the minor axes of the
substructures and the major axes of their host halos from the
physical principles,  under the assumption that the substructure
alignment on galaxy scale is a consequence of the tidal fields of
the host halo gravitational potential. Numerically, we use a
sample of four cluster-scale halos and their galaxy-scale
substructures from recent high-resolution N-body simulations to
measure the probability density distribution.  We compare the
numerical distribution with the analytic prediction, and find that
the two results agree with each other very well. We conclude that 
our analytic model provides a quantitative physical explanation 
for the intrinsic alignment of dark halo substructures. 
We also discuss the possibility of discriminating our model from the 
anisotropic infall scenario by testing it against very large N-body 
simulations in the future.
\end{abstract}

\section{INTRODUCTION}

The dark halo substructure has recently come to one of the most lively topics
in cosmology. Although the standard cosmological paradigm based on the
cold dark matter (CDM) concept generically predicts the presence of the
substructure inside the dark matter halos, there are many questions yet
to be answered associated with the dark halo substructures. The intrinsic
alignment effect of the dark halo substructure is one of those questions.

There are plenty of observational evidences that the major axes of the
brightest cluster galaxies (BCGs) have a strong tendency to be aligned with
that of their host clusters \citep{sas68,car-met80,bin82,str-pee85,rhe-kat87,
wes89,wes94,pli94,ful-etal99,kim-etal02}.  The most popular theory for the
BCG alignment is the anisotropic infall scenario based on the standard
hierarchical clustering model \citep{wes89}: The initial density field
of CDM is web-like, interconnected by the primordial filaments
\citep{bon87,bon-etal96}. The gravitational collapse and merging to form
structures occurs not in an isotropic way but in an anisotropic way along
the large-scale filaments. Accordingly, the infall of materials into a cluster
also occurs along the primordial filament, which will induce the alignment
between the orientation of a host cluster and that of BCG embedded in it.

There are several reasons that the anisotropic infall theory became so
popular: Being simple and intuitive, it fits very well into the cold dark
matter paradigm.  In addition, it has been supported by several numerical
simulations \citep[e.g.,][]{wes-etal91,van-van93,dub98,fal-etal02} which
demonstrated that the gravitational infall and merging of materials indeed
occurs along the filaments.

Nevertheless, the theory is only qualitative and still incomplete. 
Recent observations indicate that not only the BCGs but also the less 
dominant cluster galaxies exhibit the alignment effect to a 
non-negligible degree \citep{pli-bas02,pli-etal03,per-kuh04}.
In the anisotropic infall model, the substructure alignment is a primordial
effect, and would get damped away quickly by the subsequent
nonlinear processes such as the violent relaxation, the secondary infall,
and so on \citep{qui-bin92,cou96}.  Therefore, it is very unlikely that the
cluster galaxies other than the BCGs keep the primordial alignment effect 
till the present epoch \citep{pli-etal03}. Here, we propose that the
initial tidal interaction between the subhalos and the host halo will be
responsible for the observed intrinsic alignment of the cluster galaxies.

\section{ANALYTICAL PREDICTIONS}

When a subhalo forms inside a host halo, it acquires the angular
momentum ${\bf L}=(L_{i})$ due to the tidal shear field ${\bf
T}=(T_{ij})$ generated by the gravitational potential of the host
halo $(\Psi)$: $T_{ij} \equiv \partial_{i}\partial_{j}\Psi$.
\citet{lee-pen00,lee-pen01} proposed the following formula to
quantify the mutual correlations between ${\bf T}$ and ${\bf L}$:
\begin{equation}
\label{eqn:spin}
\langle L_{i}L_{j} | \hat{\bf T}\rangle = \frac{1+c}{3}\delta_{ij} -
c\hat{T}_{ik}\hat{T}_{kj}.
\end{equation}
where $c \in [0,1]$ is a correlation parameter to quantify the strength of
the correlation between $\hat{\bf T}$ and ${\bf L}$, and
$\hat{\bf T} =(\hat{T_{ij}})$ is a unit traceless tidal shear tensor defined
as  $\hat{T}_{ij} \equiv \tilde{T}_{ij}/\vert\tilde{\bf T}\vert$ with
$ \tilde{T}_{ij} \equiv T_{ij} - {\rm Tr}({\bf T})\delta_{ij}/3$, and
${\bf L} = (L_{i})$ is a rescaled but not a unit angular momentum.
If we replace the rescaled angular momentum by the unit angular momentum,
$\hat{\bf L}_{i} \equiv {\bf L}/|{\bf L}|$ in equation (\ref{eqn:spin}),
then the correlation parameter $c$ is reduced by a factor of $3/5$.
Note here that the LHS of equation (\ref{eqn:spin}) represents a
{\it conditional} ensemble average of $L_{i}L_{j}$ provided that the
unit traceless tidal shear tensor is given as $\hat{T}_{ij}$.
For the detailed explanations of equation (\ref{eqn:spin}),
see Appendix A in \citet{lee-pen01}.

It is naturally expected that $c$ decreases with time as the correlation 
between $\hat{\bf T}$ and ${\bf L}$ must decrease after the moment of the 
turn-around due to the subsequent nonlinear process. \citet{lee-pen02} found 
$c \sim 0.3$ at present epoch by analyzing the data from the Tully Galaxy 
Catalog and the Point Source Redshift Catalog Redshift Survey (in their 
original work, they used a reduced correlation parameter
$a \equiv 3c/5$ and found $a =0.18$).

Strictly speaking, equation (\ref{eqn:spin}) holds only if $\hat{\bf T}$ and
${\bf L}$ are defined at the same positions \citep{lee-pen00,lee-pen01}.
Here, they don't: $\hat{\bf T}$ and ${\bf L}$ are defined at the centers of
the mass of the host halo and the subhalo, respectively.  For simplicity,
here we just assume that  equation (\ref{eqn:spin}) still holds, ignoring
the separation between the centers of the mass of the subhalos and that
of the host halo.

The distribution of ${\bf L}$ under the influence of the tidal field is
often regarded as Gaussian \citep{cat-the96,lee-pen01}:
\begin{equation}
\label{eqn:ldis}
P({\bf L}) = \frac{1}{[(2\pi)^3 {\rm det}(M)]^{1/2}}
\exp\left[-\frac{L_{i}(M^{-1})_{ij}L_{j}}{2}\right],
\end{equation}
where the covariance matrix
$M_{ij} \equiv \langle L_{i}L_{j}| \hat{\bf T}\rangle$ is related to the
tidal shear field $\hat{\bf T}$ by equation (\ref{eqn:spin}).
In the principal axis frame of  $\hat{\bf T}$,  let us express ${\bf L}$
in terms of the spherical coordinates:
${\bf L} = (L\sin\theta\cos\phi,L\sin\theta\sin\phi,L\cos\theta)$
where $L \equiv |{\bf L}|$, and $\theta$ and $\phi$ are the polar and
the azimuthal angles of ${\bf L}$. Note that the polar angle $\theta$
represents the angle between the direction of ${\bf L}$ of the subhalo
and the minor principal axis of $\hat{\bf T}$ of its host.

The probability density distribution of the cosines of the polar angle
$\theta$ can be obtained by integrating out equation (\ref{eqn:ldis})
over $L$ and $\phi$ \citep{lee04}:
\begin{eqnarray}
p(\cos\theta) &=& \frac{1}{2\pi}\prod_{i=1}^{3}
\left(1+c-3c\hat{\lambda}^{2}_{i}\right)^{-\frac{1}{2}}\times \nonumber \\
&&\int_{0}^{2\pi}
\left(\frac{\sin^{2}\theta\cos^{2}\phi}{1+c-3c\hat{\lambda}^{2}_{1}} +
\frac{\sin^{2}\theta\sin^{2}\phi}{1+c-3c\hat{\lambda}^{2}_{2}} +
\frac{\cos^{2}\theta}
{1+c-3c\hat{\lambda}^{2}_{3}}\right)^{-\frac{3}{2}}d\phi.
\label{eqn:vtdis}
\end{eqnarray}
Here the polar angle $\theta$ is forced to be in the range of $[0,\pi/2]$
satisfying $\int_{0}^{\pi/2}p(\theta)\sin\theta d\theta = 1$, since we
care about the relative spatial orientation of the subhalo axis, but not
its sign.  Here the three $\hat{\lambda}_{i}$'s ($i=1,2,3$)
are the eigenvalues of $\hat{\bf T}$ in a decreasing order
satisfying the following two conditions: (i) $\sum_{i}\hat{\lambda}_{i}=0$;
(ii) $\sum_{i}\hat{\lambda}^{2}_{i}=1$.  If ${\bf T}$ is a Gaussian random
field which is true in the linear regime,  one can show that
$\hat{\lambda}_{1} \approx -\hat{\lambda}_{3} \approx 1/\sqrt{2}$ and
$\hat{\lambda}_{2}  \approx 0$ \citep{lee-pen01}.

We adopt the following two assumptions: (i) On cluster scale, the
principal axes of the inertia shape tensor $I_{ij}$ of a host halo is
aligned with its tidal shear tensor  $T_{ij}$ with the eigenvalues being in
an opposite order. In other words, the major principal axis of $I_{ij}$ is
the minor principal axis of $T_{ij}$, and vice versa. Note that in
\citet{lee04}, it was erroneously stated that it the major axis of $I_{ij}$
is the major axis of $T_{ij}$ (Trujillo, Carretero, \& Juncosa 2004 in
private communication); (ii) The minor axis of a subhalo is in the direction
of its angular momentum.

A justification of the first assumption is given by the Zel'dovich
approximation \citep{zel70} which predicts a prefect alignment between
the principal axis of $I_{ij}$ and that of $T_{ij}$.
Since the cluster-size halos are believed to be in the quasi-linear regime
where the Zel'dovich approximation is valid, the first assumption should
provide a good approximation to the reality.
Moreover, recent N-body simulations indeed demonstrated that  $I_{ij}$ and
$T_{ij}$ are quite strongly correlated \citep{lee-pen00,por-etal02}.
Regarding the second assumption, there is an established theory that the
spin axis of an ellipsoidal object in the gravitational tidal field is well
correlated with its minor axis \citep{bin-tre87}, which was also
confirmed by several N-body simulations \citep[e.g.,][]{fal-etal02}

Now that the minor principal axis of $\hat{\bf T}$ is the major axis of the
host halo, and ${\bf L}$ is aligned with the minor axis of the subhalo,
the polar angle $\theta$ in equation (\ref{eqn:vtdis}) actually equals
{\it the angle between the minor axis of the subhalo and the major
axis of the host halo}.
Putting $\hat{\lambda}_{1}=1/\sqrt{2}$, $\hat{\lambda}_{2}=0$, and
$\hat{\lambda}_{3}=-1/\sqrt{2}$, we simplify equation (\ref{eqn:vtdis})
into
\begin{equation}
\label{eqn:tdis}
p(\cos\theta) = \frac{1}{2\pi}(1+c)\sqrt{1 - \frac{c}{2}}
\int_{0}^{2\pi}\left[1 + c\left(1 - \frac{3}{2}\sin^{2}\theta\sin^{2}\phi
\right)\right]^{-3/2}d\phi,
\end{equation}

In the asymptotic limit of $c \ll 1$,  equation (\ref{eqn:tdis}) can
be further simplified into the following closed form:
\begin{equation}
\label{eqn:tdisa}
p(\cos\theta) = \left(1 - \frac{3c}{4}\right) +
\frac{9c}{8}\sin^{2}\theta
\end{equation}

Equations (\ref{eqn:tdis}) and (\ref{eqn:tdisa}) imply that $p(\theta)$
increases as $\theta$ increases. That is, the minor axis of a subhalo
has a strong propensity to be {\it anti-aligned} with the major axis of its
host halo.  Hence, it explains the observed alignment
effect between the subhalo and the host halo major axes as a consequence
of the intrinsic anti-alignment between the subhalo minor axis and the
host halo major axis.

The value of $c$ in equations (\ref{eqn:tdis}) and (\ref{eqn:tdisa}) should 
depend on the distance from the host halo center ($r$), the subhalo mass 
($m$) and redshift ($z$): $c = c(r,m,z)$. What one can naturally expect is 
that $c$ should decrease with $r$ since the tidal interaction must be 
strongest in the inner part of the host halo, and that $c$ should increase 
with $m$ and $r$ since the alignment effect gets reduced in the nonlinear 
regime. Unfortunately, it will be very difficult to find thefunctional form 
of $c(r,z,m)$ as $c$ contains all the nonlinear informations of galaxy 
evolution.  We do not attempt to find $c(r, z,m)$ here since it is beyond 
the scope of this Letter. Instead, we simplely assume that $c$ is a constant, 
and determine the value of $c$ empirically by fitting equation 
(\ref{eqn:tdis}) to the numerical results in $\S 3$.

\section{NUMERICAL EVIDENCES}

The data we use in this Letter is the high resolution halo simulations of
\citet{jin-sut00}. First they selected dark matter halos from their previous
cosmological P$^{3}$M N-body simulations with $256^{3}$ particles in a
$100h^{-1}$Mpc cube \citep{jin-sut98}. The halos were identified using the
standard friend-of-friend (FOF) algorithm, among which  four halos on 
cluster-mass scales (with mass around $5-10 \times 10^{14}h^{-1}M_{\odot}$) 
were then re-simulated using the nested-grid P$^{3}$M code which was designed 
to simulate high-resolution halos. The force resolution is typically $0.4\%$ 
of the virial radius, and each halo is represented by about $2 \times 10^{6}$ 
particles within the virial radius.
We then use the {\tt SUBFIND} routine of \citet{spr-etal01} to identify the
disjoint self-bound subhalos within these halos, and include those subhalos
containing more than 10 particles in the analysis.  These simulations adopted
the ``concordance'' $\Lambda$CDM cosmology with $\Omega_{0}=0.3$,
$\Omega_{\Lambda,0}=0.7$, and $h=0.7$.

Using this numerical data, we first compute the inertia tensors as
$I_{ij} \equiv \Sigma_{\alpha} m_{\alpha}x_{\alpha,i}x_{\alpha,j}$
for each host halo and its subhalos in their respective center-of-mass
frames. Then, we find the directions of the eigenvectors corresponding
to the largest and the smallest eigenvalues of the host halo and the subhalo
inertia tensors, respectively, by rotating the inertia tensors into
the principal axes frame, and determine the major axes of each host halo and
the minor axis of its subhalos. Then, we measure the cosines of the angles,
$\theta$, between the major axis of the host halo and the minor axes of their
subhalos by computing $\cos\theta \equiv \hat{\bf e}_{h}\cdot \hat{\bf e}_{s}$
where $\hat{\bf e}_{h}$ and $\hat{\bf e}_{s}$ represent the major and the
minor axes of the host halo and its subhalos, respectively.  Finally, we find
the probability density distribution, $p(\cos\theta)$, by counting the
number density of the subhalos.  When computing the probability density
distribution, we use all the subhalos in the host halo linked by the
FOF algorithm.

\begin{figure}
\begin{center}
\plotone{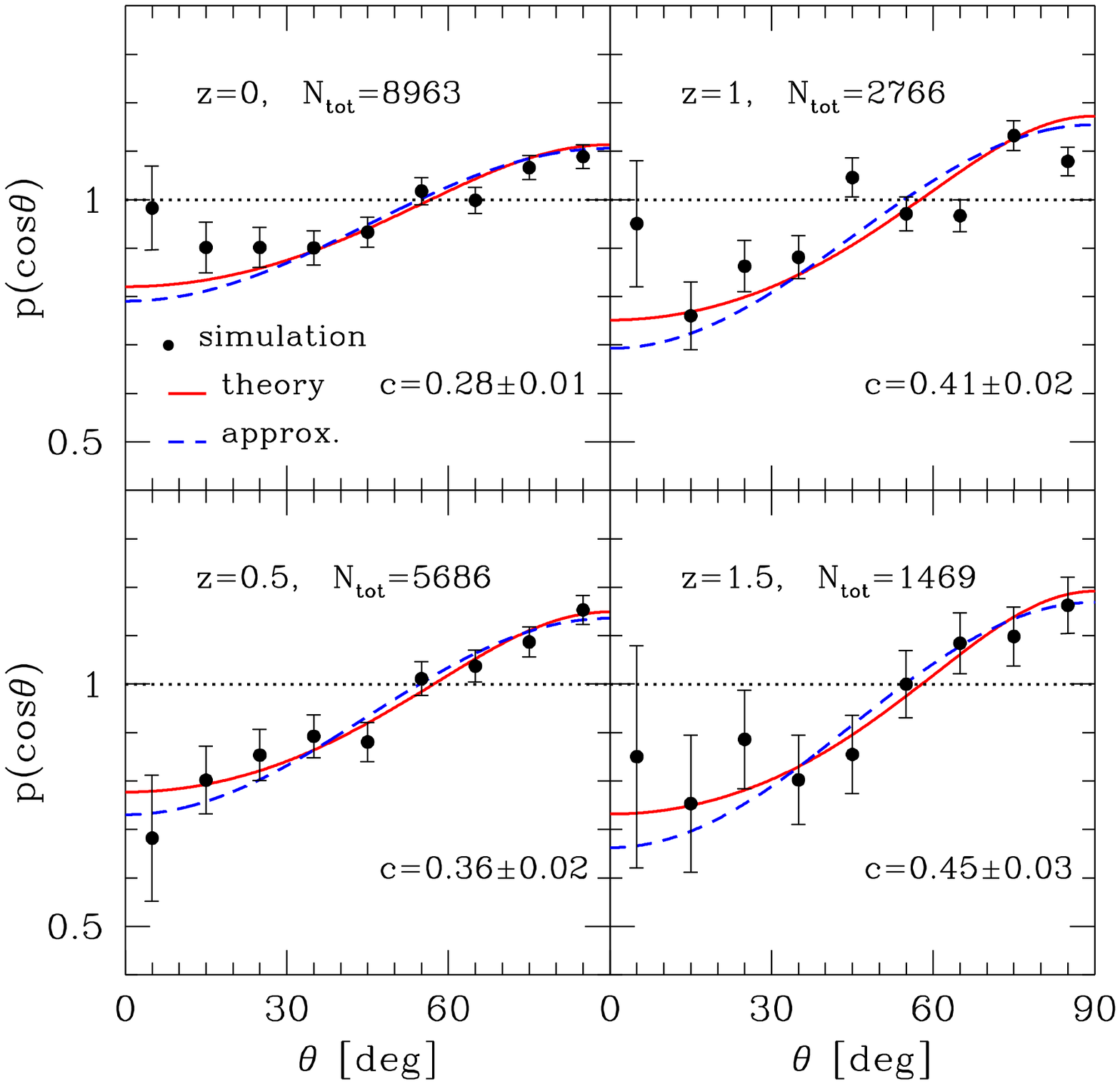} \caption{Probability density distributions of the
angles between the minor axes of the subhalo and the host halo at
four different epochs ; $z=0,0.5,1$ and $1.5$. In each panel, the
dots represent the simulation results with the Poisson errors,
while the solid and dashed lines represent the theoretical
prediction (\ref{eqn:tdis}) and the approximation formula
(\ref{eqn:tdisa}), respectively. The horizontal dotted line 
corresponds to the case of no alignment effect. \label{fig:dis1}}
\end{center}
\end{figure}

We perform the above procedure at four different redshifts:
$z=0,0.5,1$ and $2$. The total number of the subhalos $N_{tot}$ at each
redshift is $8963$, $5686$, $2766$, and $1469$, respectively. Figure
\ref{fig:dis1} plots the final numerical distributions (solid dots) with
the error bars. The error bar at each bin is nothing but the Poisson mean
for the case of no alignment given as given as $1/\sqrt{N_{bin}-1}$  
where $N_{bin}$ is the number of the subhalos at each bin.
As one can see, the numerical distribution $p(\cos\theta)$ increases as 
$\theta$ increases, revealing that the minor axes of substructures really 
tend to be anti-aligned with the major axes of their host halos, as 
predicted by the analytic model (eq.[\ref{eqn:tdis}]) of $\S 2$.
Figure \ref{fig:dis1} also plots the analytic predictions (solid line) and
the approximation formula (dashed line) derived in $\S 2$. The horizontal 
dotted line represents the uniform distribution of $\cos\theta$ for the 
case of no alignment. 

We fit the analytic distributions to the numerical data points to determine
the best-fit values of the correlation parameter $c$.
We find $c = 0.28\pm 0.01, 0.36\pm 0.02, 0.41\pm 0.02, 0.45\pm 0.03$ at
$z=0,0.5,1,1.5$, respectively. The errors involved in the determination of
$c$ is given as the standard deviation of $c$ for the case of no alignment
effect given as $\epsilon_{c} \equiv \sqrt{c^{2}/(N_{tot}-1)}$ where 
$N_{tot}$ is the number of all subhalos used to compute $c$.
The value of $c$ increases with redshifts $z$, as expected.

In fact the value of $c=0.3$ gives quite a good fit, if not the best, 
not only at the present epoch of $z=0$  but also at all earlier
epochs of $z=0.5,1,1.5$, which implies that the initially induced
anti-alignment effect is more or less conserved, reflecting the fact
that the directions of the subhalo angular momentum are fairly well conserved.

\begin{figure}
\begin{center}
\plotone{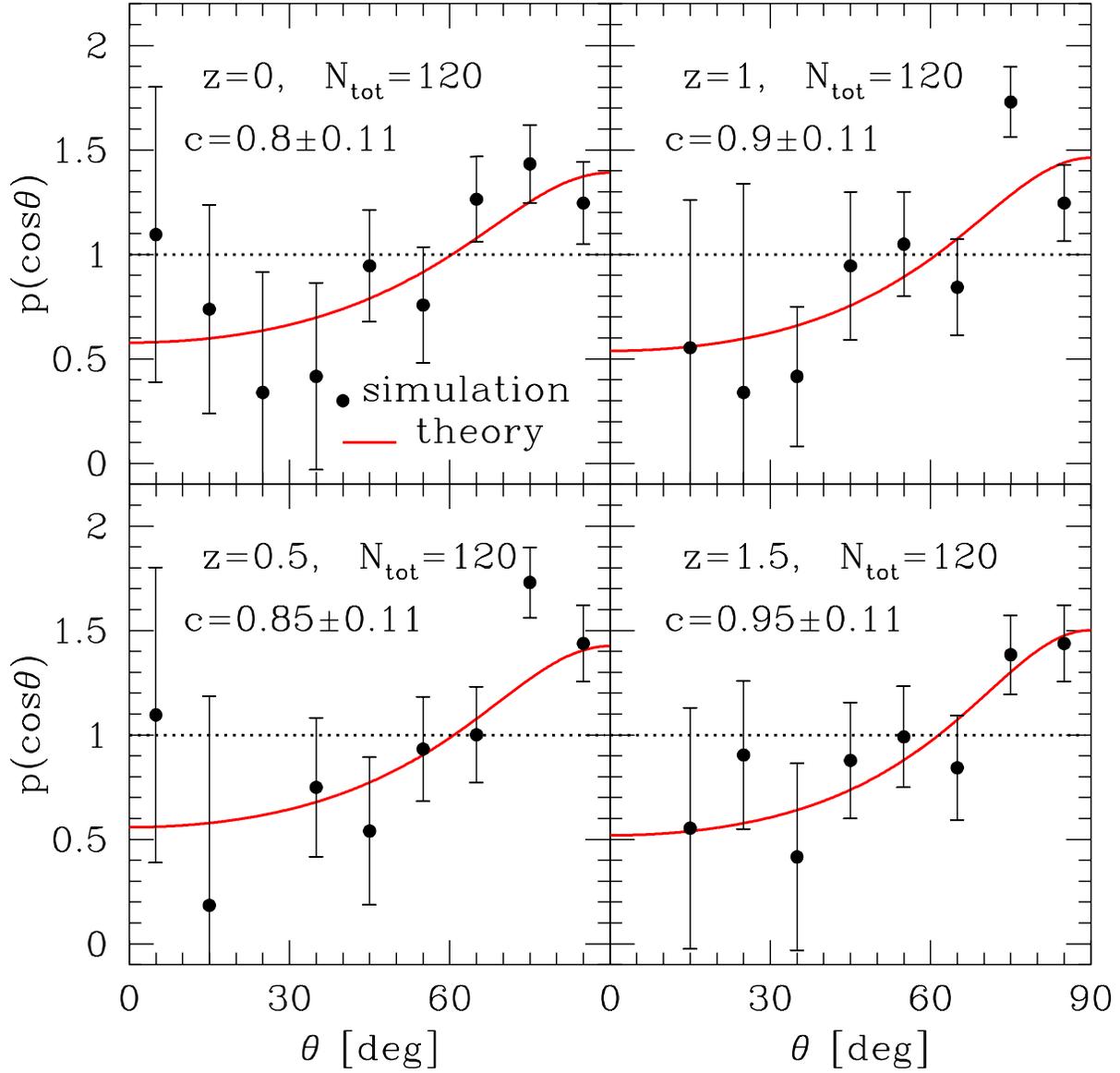}
\caption{Same as figure \ref{fig:dis1} but with the most massive
30 subhalos.
\label{fig:dis2}}
\end{center}
\end{figure}

To understand the dependence of the alignment effect on the subhalo mass, we
derive the same probability distribution but by using only the most massive
30 subhalos in each cluster ($N_{tot}=120$ for each redshift), and find the
corresponding best-fit values of $c$.  We find $c=0.8\pm 0.11,
0.85\pm 0.11,0.9\pm 0.11,0.95\pm 0.11$ at $z=0,0.5,1,1.5$, respectively.
Figure \ref{fig:dis2} plots the results. The approximation formula
(eq.[\ref{eqn:tdisa}]) is excluded in this figure since the best-fit values
of $c$ for this case is pretty close to unity. Although the large error bars
prevent us from making a quantitative statement, it is obvious that the
anti-alignment effect is stronger for the case massive subhalos.

\section{SUMMARY AND DISCUSSIONS}

Although the currently popular anisotropic merging and infall scenario has
provided a qualitative explanation for the BCG-cluster and
cluster-cluster alignments \citep[e.g.,][]{hop-etal05}, no previous approach
based on this scenario was capable of making a quantitative prediction for
the alignment effect of cluster galaxies other than BCGs with the host
clusters. We constructed an analytic model where the substructure alignment 
is a consequence of the tidal field of the host halo at least
on the scale of cluster galaxies, and predicted quantitatively the strength
of the alignment effect, by comparing the model with the results from recent
high-resolution N-body simulations.

However, it is worth noting that we have yet to completely rule out the 
anisotropic infall model.  An idealistic way to discriminate our analytic 
model from the anisotropic infall scenario would be to measure directly the
correlation of the directions of the subhalo angular momentum with the host
halo orientations. Unfortunately, it is very difficult to determine 
the directions of the subhalo angular momentum vectors in current simulations.
Because the rotation speed of a dark halo in simulations is only a few percent 
of the virial motion of the particles, one needs more than $10^4$ particles 
to determine accurately the direction of the angular momentum vector of a 
subhalo with an average rotation speed. In current simulations the subhalos 
have much fewer particles than $10^4$.  This is why we used the minor axes 
of the subhalos rather than the directions of the subhalo angular momentum 
vectors to investigate the intrinsic alignments of substructures.

Nevertheless, the strong alignment between the halo minor axes and angular 
momentum vectors \citep{bin-tre87} should indicate indirectly that the 
anti-alignments between the subhalo minor axes and the host halo major axes 
observed in our simulations are likely to be caused by the host halo tidal 
field as our model predicts.  
Many N-body simulations \citep{bar-efs87,dub92,war-etal92,bai-etal05} have 
already proved that the dark matter halos rotate and their angular momentum 
vectors are aligned with their minor axes. Furthermore, \citet{bai-ste04} 
demonstrated evidently that there are good internal alignments between the 
halo minor axes and angular momentum vectors measured at different radii. 
Therefore, the alignments between the halo angular momentum vectors and 
the minor axes are expected to be hold even when the outer parts of the 
halos get disrupted when they fall into larger halos as substructures. 
Indeed,  we ourselves check this effect in our simulations: we measure 
the angular momentum vectors of several very massive subhalos with more than 
$10^{4}$ particles, and find that the subhalos have {\it non-zero} angular 
momentum and that the subhalo minor axes are strongly aligned the directions 
of their angular momentum: the cosines of all alignment angles turn 
out to be bigger than $0.6$. 

Anyway, it will be definitely necessary to investigate directly the
correlations of the directions of the subhalo angular momentum with 
the orientations of their host halos in the future with larger 
simulation data, where the number of particles belonging to subhalos 
should be large enough. Using larger simulation data, it will be 
also possible to determine the functional form of the correlation 
parameter $c(r,m,z)$. Our future work is in thisdirection.

\acknowledgments

We thank the anonymous referee who helped us improve the original manuscript.
J.L. wishes to thank the Shanghai Astronomical Observatory for a warm
hospitality during the workshop on Cosmology and Galaxy Formation where
this collaboration began.  J.L. is supported by the research grant
of the Korea Institute for Advanced Study.  X.K. and Y.P.J. are partly 
supported by NKBRSF (G19990754), by NSFC(Nos. 10125314, 10373012), 
and by Shanghai Key Projects in Basic Research (No. 04jc14079)

\clearpage

\clearpage

\end{document}